\documentclass[aps,amsmath,prb,a4paper,showpacs,twocolumn]{revtex4}
\usepackage{graphicx}

\newcommand{\eqn}[1]{(\ref{#1})}
\newcommand{\beq}{\begin{equation}}
\newcommand{\eneq}{\end{equation}}
\newcommand{\bea}{\begin{eqnarray}}
\newcommand{\enea}{\end{eqnarray}}
\newcommand{\bean}{\begin{eqnarray*}}
\newcommand{\eean}{\end{eqnarray*}}

\newcommand{\comment}[1]{}
\newcommand{\R}{\mathrm{R}}
\newcommand{\Le}{\mathrm{L}}
\newcommand{\fract}[2]{\frac{\displaystyle #1}{\displaystyle #2}}
\newcommand{\dagga}{{\phantom{\dagger}}}

\begin{document}

\title{Two-level Physics in a Model Metallic Break Junction}
\author{P. Lucignano$^{1,2}$}
\author{G.E. Santoro$^{1,3}$}
\author{M. Fabrizio$^{1,3}$}
\author{E. Tosatti$^{1,3}$}

\affiliation{$^1$ SISSA and CNR-INFM Democritos National Simulation Center,
             Via Beirut 2-4, 34014 Trieste, Italy}

 \affiliation{$^2$
Coherentia CNR-INFM and Dipartimento di Scienze Fisiche Universit\`a di
Napoli, "Federico II", Monte S.Angelo - via Cintia, I-80126 Napoli, Italy.}

\affiliation{$^3$ International Centre for Theoretical Physics (ICTP), 
             P.O. Box 586, I-34014 Trieste, Italy}
\begin{abstract}

We consider a model inspired by a metal break-junction hypothetically caught at its breaking 
point, where the non-adiabatic center-of-mass motion of the bridging atom can be treated
as a two-level system. By means of Numerical Renormalization Group (NRG) we calculate the influence
of the two level system on the ballistic conductance across the bridge atom. The results 
are shown to be fully consistent with a conformal field theory treatment.  We find that the 
conductance, calculated by coupling Fermi liquid theory to our NRG is always 
finite and fractional at zero temperature, but drops quite fast as the temperature increases.

\end{abstract}

\maketitle

\section{Introduction}

A number of transport measurements on organic and inorganic molecules     
bridged bewteen metallic leads have recently succeeded in revealing signatures of 
the molecular vibrational and motional degrees of freedom in the inelastic 
tunneling spectrum, and raised interesting theoretical issues. Most notably, 
since in these nanosized devices the time scales of the nuclear dynamics may 
be comparable to those involved in the electron tunneling, non-adiabatic quantum 
effects become not negligible. This question has been the subject of extensive 
theoretical activity over the past years, mostly concerned with the 
vibrational effects, for which we refer to a recent review~\cite{galperin} and 
to the references therein. The role of the center-of-mass oscillations of a bridging
site between the two leads has been well addressed, mainly via generalized Master equations
in the context of nanoelectromechanical quantum-shuttle devices. On the contrary, 
the low-temperature quantum-coherent regime has been only slightly touched, 
and with rather controversial results at that. For instance, Al-Hassanieh 
{\sl et al.}~\cite{Dagotto} made use of exact diagonalization procedure
supplemented by a Dyson-equation embedding to conclude that conductance should
be suppressed in resonance conditions for arbitrary coupling strength between 
the center-of-mass motion and the hybridization with the leads, and both 
at finite and vanishing charging energy. This result was questioned by Mravlje 
{\sl et al.}~\cite{Mravlje} who found, by a variational procedure and for finite 
charging energy, that the center-of-mass motion does not affect perfect 
transmission at resonance.   

\begin{figure}[!htp] 
\includegraphics*[width=0.7\linewidth]{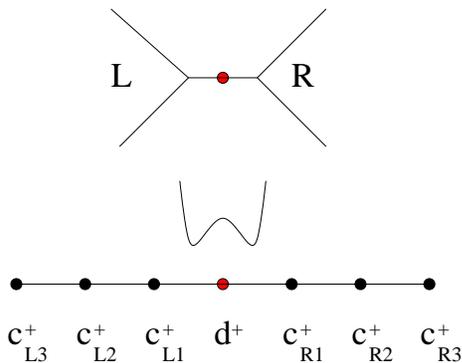} 
\caption{Sketch of a mechanical break junction} \label{fig1}
\end{figure}

In this paper we address the same class of questions, concerning the role 
of the center-of-mass motion at low temperature, in a different type of 
systems, namely metallic break junctions (BJ).~\cite{break_junction} 
In a BJ the metal bridge or neck, initially forming a single solid body
strongly bonded with the leads, is mechanically broken apart typically
at criogenic temperatures. The conductance drops prior to breaking typically
takes place through a sequence of plateaus corresponding to thinning of the
neck, down the ultimate monatomic contact, whose conductance 
is of the order of the conductance quantum $G_0 = 2e^2/h$, where e and h are
the electron charge and Planck's constant. These plateaus,
are interpreted in terms of ballistic conductance, which in the adiabatic 
Landauer-Buettiker linear response theory\cite{LB1,LB2} is controlled 
by the few residual one-electron conduction channels and by their respective 
transmittivity. The instant when the left and the right leads are separating,
the physical bridge between the two is as a rule a single metal atom --
as indicated by the last conductance plateau.\cite{break_junction} 
Here, non-adiabatic effects could in principle be caught right at the moment of separation.
The bridge atom, initially strongly attached to both leads, eventually detaches from 
one of them to remain after separation exclusively attached to the other. 
In the process, the atom coordinate will move for a while in a double well 
effective potential. Therefore, between the initial solid
metal-metal nanocontact, held together by a strong bond and with electrical
properties governed by ordinary ballistic conductance, and the broken contact,
there is room for a transient state where a new regime involving double well tunneling 
may be relevant. As the double well initially develops out of a 
flat single well, the two well minima can to a good accuracy be considered 
equivalent; moreover the barrier separating them is initially very weak, 
which calls for quantum tunneling, even when the atom mass is not small.    
A skematic sketch of this system is shown in Fig.~\ref{fig1}.
If the mechanical breaking takes place slowly enough in time, the dynamics of the bridge atom 
nucleus tunneling in the double well may be approximated by that of a two-level system (TLS), whereas
the electronic level of the bridge atom, assumed to be nondegenerate, 
gives rise to a resonant electronic level.

Within these assumptions, the physics might be assimilated to that 
of conduction electrons scattering off TLS's in bulk metals, proposed  
by Vladar and Zawadowski~\cite{zawadowski} as a possible realization of 
a two channel Kondo (2CK) model.~\cite{nozieres} This idea recurred
several times in recent years in the context of a variety phenomena in 
metals~\cite{ralph,halbritter,cichorek,zawadowski2,kolesnychenko,cox}, although  
again rather controversially. According to Aleiner {\sl et al.}~\cite{altshuler} 
in fact, the appropriate high-energy cut-off of a TLS coupled to conduction 
electrons is not the electron bandwidth but rather the oscillation 
frequency within each potential well, since above this energy 
the conduction electrons follow adiabatically the motion of the atom. 
Since the Kondo temperature, $T_K$, is typically 
exponentially smaller than the high-energy cut-off, the 
conclusion of Aleiner {\sl et al.} is that $T_K$ is too small to be relevant.

The BJ problem proposes now a new interesting physical situation which
we treat here in a slightly different model, arriving at interesting conclusions
about the zero temperature conductance and its temperature evolution.

By means of the Numerical Renormalization Group (NRG),~\cite{wilson}  we calculate the influence
of the two level system on the ballistic conductance across the bridge atom.
We find that the zero temperature, zero voltage conductance is always finite 
and fractional. However it is found to drop quite fast to zero as the temperature increases.

The paper is organized as follows.
In Section II we introduce the model Hamiltonian and discuss the parameters chosen.
In Section III we first solve some limiting cases by mean of analytical methods.
In Section IV we study the low energy properties of our model by mean of NRG.
We show that conformal field theory (CFT) provides a strikingly direct interpretation of 
the low lying spectrum obtained by NRG. In Section V by using both our NRG  routine 
and a Fermi liquid theory we give an estimate of the conductance of our model.
In Section VI we finally summarize and comment our results.

\section{Model Hamiltonian}  

The physics and language of our model is inspired by a bridge atom suspended 
between two one-dimensional metallic leads and moving quantum-mechanically in a symmetrical 
double well potential, although the model could equally other pseudospin variables coupled to 
a ballistic conductance channel. As a simplification we will assume that the dynamics 
of the atom nuclear coordinate is that of a TLS. We introduce a pseudo-spin 
variable $\tau_z$  identifying the atom position, $\tau_z=1$ and $\tau_z=-1$ 
when the atom is in the minumum close to the right (R) and left (L) lead respectively. 
With this definition, assuming the bridge atom wavefunction to be 
real~\cite{zawadowski}, the quantum tunneling operator between the two 
wells corresponds to the Pauli matrix $\tau_x$. 

The right and left metal leads are modeled as semi-infinite chains, site-label $n=1,\dots,\infty$, 
with nearest neighbor hopping, amplitude $-t$, 
and creation (annihilation) operators $c^\dagger_{\alpha\,n\sigma}$ ($c^\dagga_{\alpha\,n\sigma}$), 
where $\alpha=\R,\Le$ and the spin $\sigma=\uparrow,\downarrow$.  The bridge atom is
endowed with a single nondegenerate electronic orbital (the bridge level),
of creation and annihilation operators 
$d^\dagger_\sigma$ and  $d^\dagga_\sigma$,respectively, constituting the
ballistic conducting channel. The 
electron hopping 
amplitude from the leads to the bridge level is assumed 
to depend explicitly on $\tau_z$.  When the atom is in the right well ($\tau_z=+1$), 
the level is more coupled to 
the R chain, amplitude $-t_0(1+\gamma)$ with $0\leq \gamma\leq 1$, than to the L chain, amplitude $-t_0(1-\gamma)$, 
and viceversa when the atom is in the left well ($\tau_z=-1$). Therefore the model Hamiltonian reads

\bea
\mathcal{H_0} &=& -t\,\sum_{\alpha=\R,\Le}\,\sum_\sigma\,\sum_{n=1}^\infty\, 
c^\dagger_{\alpha\,n\sigma}c^\dagga_{\alpha\,n+1\sigma} + H.c.\nonumber\\
&& -t_0\,\sum_\sigma\, \Big(1+\gamma\,\tau_z\Big)\,\Big(c^\dagger_{\R\,1\sigma}d^\dagga_\sigma + H.c.\Big)\nonumber\\
&& -t_0\,\sum_\sigma\, \Big(1-\gamma\,\tau_z\Big)\,\Big(c^\dagger_{\Le\,1\sigma}d^\dagga_\sigma + H.c.\Big)\nonumber\\
&& - \Delta_x\,\tau_x - V_x\,\tau_x\,\sum_\sigma\,d^\dagger_\sigma d^\dagga_\sigma.
\label{Ham-1}
\enea 

The last term represents the electron assisted tunneling of the bridge atome 
nucleus arising from the influence of the atom's state of charge on the height 
of the barrier of the double well tunnelling potential~\cite{zawadowski}.
In principle this type of assisted tunneling process includes other possible 
operators that couple the bridge level and the nuclear pseudospin coordinate, 
provided (given our assumption of a symmetric double well and equivalent leads) 
they are equally even under under reflection with respect to the center of the 
double well (we will call this even parity).
The last term in \eqn{Ham-1} is therefore just one of the operators that presumably 
might possess a large matrix element, involving the bridge level charge occupancy. 
In later calculations below we will actually consider more general assisted tunneling 
operators too. 

One can note at the outset that the model in \eqn{Ham-1} is closely related to a 2CK 
model, the role of the spin being played by the lead label, R and L, for the 
conduction electrons and by the pseudospin $\vec{\tau}$ that identifies the TLS, 
while the role of the silent channels is played by the real spin $\sigma$. 
An alternative way of writing \eqn{Ham-1}, which may be convenient in some cases, 
is by introducing the even ($e$) and odd ($o$) combinations

\begin{eqnarray}
c^\dagga_{e\,n+1\sigma} &=& \sqrt{\frac{1}{2}}\left(c^\dagga_{\R\,n\sigma} + c^\dagga_{\Le\,n\sigma}\right),\label{fermi_rot1}\\
c^\dagga_{o\,n\sigma} &=& \sqrt{\frac{1}{2}}\left(c^\dagga_{\R\,n\sigma} - c^\dagga_{\Le\,n\sigma}\right),\label{fermi_rot2}
\end{eqnarray}
and formally defining 
\[
c^\dagga_{e\,1\sigma} = d_\sigma,
\]

through which the model \eqn{Ham-1} is rewritten as 

\bea
\mathcal{H} &=& -t\,\sum_{\alpha=e,o}\,\sum_\sigma\,\sum_{n=1}^\infty\, 
c^\dagger_{\alpha\,n\sigma}c^\dagga_{\alpha\,n+1\sigma} + H.c.\nonumber\\
&& -\left(V_e-t\right)\,\sum_\sigma\, \Big(c^\dagger_{e\,1\sigma}c^\dagga_{e\,2\sigma} + H.c.\Big)\nonumber\\
&& -V_o\,\tau_z\,\sum_\sigma\,\Big(c^\dagger_{o\,1\sigma}c^\dagga_{e\,1\sigma} + H.c.\Big)\nonumber\\
&& - V_x\,\tau_x\,\sum_\sigma\,\left(c^\dagger_{e\,1\sigma}c^\dagga_{e\,1\sigma} - \xi\,c^\dagger_{o\,1\sigma}c^\dagga_{o\,1\sigma}
+\eta\, c^\dagger_{e\,2\sigma}c^\dagga_{e\,2\sigma}\right)\nonumber\\
&& - \Delta_x\,\tau_x, 
\label{Ham-2}
\enea 

where 

\begin{equation}
V_e = \sqrt{2}\,t_0,\qquad V_o =\sqrt{2}\,t_0\,\gamma.\label{Ve-Vo}
\end{equation}  

In $H$ of \eqn{Ham-2} we in fact included additional assisted tunneling operators with 
coupling constants parametrized by $\eta$ and $\xi$, which are missing in Eq.~(\ref{Ham-1}).
In the $even-odd$ formulation, the analogy with a 2CK model is much more explicit, 
especially once we rotate the pseudospin by $\pi/2$ around the $y$-axis, even, 
$e$, and odd, $o$, labels playing the role of spin up ($\Uparrow$) and down ($\Downarrow$).
A similar model was recently proposed by Zarand in the context of TLS's in 
metals.~\cite{zarand} 
according to whom the presence of the resonant level may push the equivalent 
2CK model into a strong coupling regime with a large Kondo temperature
of the same order as the high-energy cut-off~\cite{onzarand}.
For comparison, we may also write the conventional two-channel flavour-Kondo 
model (after a $\pi/2$ rotation around the $y$-axis of the flavour pseudo-spin)
 
\begin{eqnarray}
\mathcal{H}_{2CK}&=& -t\,\sum_{\alpha=e,o}\,\sum_\sigma\,\sum_{n=1}^\infty\, 
c^\dagger_{\alpha\,n\sigma}c^\dagga_{\alpha\,n+1\sigma} + H.c.\nonumber\\
&& + \sum_{a=x,y,z}\, J_x\,\tau_x\,T_1^z + J_y\,\tau_y\,T_1^y - J_z\,\tau_z\,T_1^x
,\label{Ham-2CK}
\end{eqnarray}

where 

\begin{equation}
T_n^a = \frac{1}{2}\sum_{\alpha\beta=e,o}\,\sum_\sigma\, c^\dagger_{\alpha\,n\sigma}\,\sigma^a_{\alpha\beta}\,
c^\dagga_{\beta\,n\sigma},
\label{flavour}
\end{equation}

are the local generators of the flavour SU(2), with $\sigma^a$ the Pauli matrices. 

Our model $H$ in \eqn{Ham-2} differs from the anisotropic 2CK model \eqn{Ham-2CK} since 

\begin{itemize}
\item the even ($\Uparrow$) chain has one more site than the odd one ($\Downarrow$); 
\item in the even ($\Uparrow$) chain the hopping between sites 1 and 2 differs from the others;
\item a local magnetic field $\Delta_x$ acts on the pseudospin.
\end{itemize}

In addition, $V_x$ in \eqn{Ham-2} is generally coupled to an operator more 
complicated than $T_1^z$, unlike $J_x$ in \eqn{Ham-2CK}. This difference has 
no effect when $\Delta_x=0$, in which case our model Eq.~\eqn{Ham-2} will display 
the conventional 2CK behavior, but plays an important role when a finite 
$\Delta_x$ drives the model away from the 2CK fixed point. Specifically, we found 
that models with different $\xi$ and and $\eta$ in Eq.~\eqn{Ham-2}, may fall 
into two different classes:

\begin{itemize}
\item[(i)] if $\eta=0$ and $\xi=1$, hence the assisted tunneling term  $V_x\,\tau_x$ 
in \eqn{Ham-2} is proportional to $T_1^z$, or, more generally, if 

\begin{equation}
\xi=1+\eta,\label{condition}
\end{equation}

then an intermediate 2CK crossover regime should survive in presence of a small but finite $\Delta_x$;
\item[ii)] if Eq.~\eqn{condition} is not satisfied, them this crossover regime 
is likely to be absent for any $\Delta_x\not=0$. In this case the model with 
$\xi=\eta=0$ can be taken as representative of all the others.  
\end{itemize}
We note that the condition \eqn{condition} means simply that the assisted tunneling operator,  
\[
\sum_\sigma\, c^\dagger_{e\,1\sigma}c^\dagga_{e\,1\sigma} - \xi\,c^\dagger_{o\,1\sigma}c^\dagga_{o\,1\sigma}
+\eta\, c^\dagger_{e\,2\sigma}c^\dagga_{e\,2\sigma},
\]

is orthogonal to the local charge density, 

\begin{eqnarray*}
&&\sum_\sigma\, c^\dagger_{e\,1\sigma}c^\dagga_{e\,1\sigma} + c^\dagger_{o\,1\sigma}c^\dagga_{o\,1\sigma}
+ c^\dagger_{e\,2\sigma}c^\dagga_{e\,2\sigma}\\
&& = \sum_\sigma \, d^\dagger_\sigma d^\dagga_\sigma + c^\dagger_{\R\,1\sigma}c^\dagga_{\R\,1\sigma}
+ c^\dagger_{\Le\,1\sigma}c^\dagga_{\Le\,1\sigma}
\end{eqnarray*}

The fact that such a property discriminates betweeen two quite distinct 
classes of behaviors suggests that the charge degrees of freedom 
play in this problem an active role, unlike in conventional Kondo models, 
as we are going to discuss in what follows.

\section{Preliminary analysis of the model}

Simplifying the double well dynamics of the bridge atom to a TLS form
permits a numerical analysis of the original model $H_0$ \eqn{Ham-1}.
We performed that analysis by means of the numerical renormalization group~\cite{wilson} 
and the results will be presented and discussed later. Prior to doing that
we can, exploiting the analogy with a 2CK problem, discuss first some 
instructive limiting cases of \eqn{Ham-1} that can be easily understood.

First, if $\Delta_x=V_x=0$ the model describes a conventional 
electron hopping
across the bridge level with inequivalent leads because of $\gamma\not = 0$. 
In particular, for any value of $\tau_z$, the zero temperature differential 
conductance in units of $G_0 = 2e^2/h$ is readily found to be \cite{datta}
 
\beq  
\fract{G}{G_0} = \frac{1-\gamma^2}{1+\gamma^2}.\label{G-allx=0}
\eneq

If $\gamma=0$ with finite $\Delta_x$ and $V_x$, 
it is more convenient to use the even-odd representation in which the 
conductance is

\beq
\fract{G}{G_0} = \sin^2\left(\delta_e-\delta_o\right),\label{G-e-o}
\eneq

where $\delta_e$ and $\delta_o$ are the phase shifts at the chemical potential 
in the even and odd channels, respectively, determined by coupling of the leads
to the bridge level. By solving the one-dimensional scattering problem 
and choosing for simplicity $\eta=\xi=0$, we find that 

\begin{eqnarray*}
\delta_e &=& \frac{\pi}{2}+\frac{tV_x}{2t_0^2}\,\tau_x,\\
\delta_o &=& 0,
\end{eqnarray*}
so that 
\beq
\fract{G}{G_0} = \fract{4t_0^4}{4t_0^4+t^2 V_x^2},\label{G-gamma=0}
\eneq

is always finite.

\subsection{Asymptotic solution for strong electron-nucleus coupling: $\gamma=1$}
    
The parameter $\gamma$ in \eqn{Ham-1} measures the strength of ``electron-phonon''
coupling between the bridge atom and the leads. When the bridge atom double well is tiny,
the two minima are close, and $\gamma$ will be small; in a wide double well, with the bridge atom very 
close to either R or L leads, $\gamma$ will be large (while $\Delta_x$ will correspondingly
be small).  The upper limit for $\gamma$ is $\gamma=1$, when the bridge atom 
in the left (right) well only couples to the left (right) lead. As it turns
out, this limit is interesting by itself. 

Since the bare electron hopping $t_0$ is of the order of the eV, which is
many orders of magnitude larger than both $\Delta_x$ and $V_x$, one can safely treat the latter terms 
perturbatively within the path-integral formalism originally developed by  
Anderson and Yuval~\cite{Anderson&Yuval} and by Hamann~\cite{Hamann} for the single-channel Kondo and 
Anderson-impurity models respectively. That approach had in turn been built by 
extending the Nozi\`eres-De Dominicis solution~\cite{Nozieres&DeDominicis} of the 
X-ray edge singularity to a succession of emission-absorption processes. 
In our problem, because of the presence of the silent spin-channel and of 
the bridge level, it is necessary to resort to a multichannel extension of 
the Anderson-Yuval formalism~\cite{vladar,mio}, where the perturbation expansion 
consists of a series of pseudo-spin flips induced by the operator $\tau_x$. 
What is important in the calculation is the 
phase-shift difference suffered by each channel at any pseudo-spin flip.
In the present case the most convenient representation is in terms of R and L leads.   
We do not present details of the calculations, since as it turns out the final result 
can be inferred by very simple arguments. 
Because as was said when $\gamma=1$ and $\tau_z=+1$, only the R lead is hybridized 
with the level, while the L lead is untouched, R acquires a phase shift $\delta^+_R=\pi/2$, 
corresponding to a resonant level model, while for the left lead, L, $\delta^+_L=0$. 
Viceversa, for $\tau_z=-1$, it is only the L lead that is coupled 
hence $\delta^-_R=0$ while $\delta^-_L=\pi/2$. Therefore the phase shift 
differences in the pseudospin flip from $\tau_z=-1$ to $\tau_z=+1$ 
are $\delta_R = \delta^+_R - \delta^-_R = \pi/2$ and $\delta_L = \delta^+_L - \delta^-_L = -\pi/2$ 
for each spin $\sigma$, which here plays the role of a silent channel. 
This is exactly the location of the so-called Emery-Kivelson point,~\cite{Emery&Kivelson} 
which also coincides  with the intermediate coupling fixed point of the 2CK model~\cite{A&L+Cox,mio}.  
Interesting enough, this situation should also correspond to the maximum Kondo 
temperature attainable,~\cite{onzarand} confirming Zarand's expectation~\cite{zarand}. 
We find that, at equilibrium, the perturbative expansion of the partition 
function coincides with that of the  generalized resonant level model 

\begin{eqnarray}
\mathcal{H}_* &=& \mathcal{H}_0\left[\psi_f,\psi_{sf}\right] - \Delta_x\,\sqrt{\frac{2v}{\Gamma}}\,
\bigg(\psi^\dagger_f(0) f + f^\dagger\,\psi^\dagga_f(0)\bigg)
\nonumber\\
&& - V_x\,\sqrt{\frac{2v}{\Gamma}}\,\bigg(f^\dagger-f\bigg)\,\bigg(\psi^\dagger_{sf}(0)+\psi^\dagga_{sf}(0)\bigg),\label{Ham-E&K}
\end{eqnarray}

where $\Gamma=4t_0^2/t$ is the hybridization width of the $d$-level, 
which plays the role of the high-energy cut-off, and 

$\mathcal{H}_0\left[\psi_f,\psi_{sf}\right]$ is the continuum limit of a non-interacting 
Hamiltonian on a closed chain for two different chiral Fermi fields 
$\psi_f(x)$ and $\psi_{sf}(x)$ that move with Fermi velocity $v$,~\cite{Emery&Kivelson} namely 

\[
\mathcal{H}_0\left[\psi_f,\psi_{sf}\right] = iv\,\sum_{a=f,sf}\,\int dx\,\psi^\dagger_a(x)\,\partial_x\psi^\dagga_a(x),
\]

and, finally, $f$ and $f^\dagger$ are the annihilation and creation operators of an auxiliary fermion 
satisfying $f^\dagger\,f - 1/2 = \tau_z$.
 
Here we labeled the fields following Emery and Kivelson~\cite{Emery&Kivelson} 
to stress the fact that the role of spin $s$ (here the real spin $\sigma$) and of flavour $f$ 
(here the R and L leads) are interchanged in our model with respect to the conventional 2CK model. 

Unlike the Emery-Kivelson Hamiltonian~\cite{Emery&Kivelson} for the 2CK model, 
in our case a pseudo-spin field $\Delta_x$ is present, which spoils the 
anomalous 2CK behavior.~\cite{pangcox} For any finite $\Delta_x$, the 
spectrum of the Hamiltonian \eqn{Ham-E&K} is Fermi-liquid like, corresponding
in fact to a 2CK model in presence of a magnetic field applied to the 
impurity, a case studied by Affleck, Ludwig, Pang and Cox.~\cite{A&L+Cox} 
We further note that the original Hamiltonian \eqn{Ham-2} is invariant under a generalized 
parity operator 

\begin{equation}
\mathcal{P} = \tau_x\,\left(-1\right)^{N_o},\label{new-parity}
\end{equation}

where $N_o$ is the total number of electrons in the odd channel. Since a 
Fermi-liquid spectrum implies that the TLS -- the Kondo impurity -- 
asymptotically dissolves into the conduction bath, it follows 
that the value on each state of the generalized parity operator 
\eqn{new-parity} turns effectively into the ``Fermi-liquid'' parity $(-1)^{N_o}$. 
This observation implies a zero-bias conductance dictated by
the form \eqn{G-e-o} in the low-energy spectrum. 

We conclude by briefly discussing the other limit $\Delta_x=0$, 
when the bridge atom is so heavy, or the barrier so large, that 
double well tunneling is suppressed. Here the model flows to the 2CK fixed point,
and here it is well known that the elastic scattering $S$-matrix at 
the chemical potential is zero.~\cite{A&L-green} Since the even and odd channels correspond in 
our model to the spin up and down channels in the 2CK problem, both have vanishing 
$S$-matrix, hence the conductance is zero. 
For an infinitesimally small magnetic field acting on the impurity spin, it was
shown in Ref.~\onlinecite{A&L+Cox} that a Fermi liquid behavior is recovered 
with a phase shift difference of $\pi/2$ between the two spin channels. 
The translation of this result in our case is not so straightforward since, 
in the absence of any coupling to the TLS, i.e. $V_o=V_x=0$, the even and odd phase 
shifts are finite unlike the conventional 2CK. Actually, since the 
even chain has one more site, the ``bare'' phase shift difference is already 
$\pi/2$. One possibility appears to be that the $\pi/2$ phase-shift difference 
acquired by switching on an infinitesimal $\Delta_x$ at the 2CK fixed point adds to the 
``bare'' value to give a total difference of $0$ modulous $\pi$. This would 
imply zero conductance for $\Delta_x\ll V_x$, raising to nonzero by 
increasing $\Delta_x$. We shall see that this is actually what happens if 
Eq.~\eqn{condition} is satisfied, namely if the assisted tunneling does not 
involve charge degrees of freedom. In the opposite case, the conductance behavior is more complicated.

\section{Numerical results}

We address the spectrum of the model Hamiltonian $H$ of \eqn{Ham-2}  by standard 
NRG~\cite{wilson}, whose results we are going to present in this section. 
Tentative values of the Hamiltonian parameters which we adopted are a 
conduction bandwidth $2D_0\sim 2$~eV, the attempt frequency $D \sim 10^{-2}$~eV,
~\cite{altshuler,zarand} $V_e\simeq V_o\sim 0.2$~eV and $V_x\simeq \Delta_x \sim 10^{-3}~V_e$. 
As discussed previously, only the conduction electrons with energy smaller than 
the attempt frequency are involved in the pseudo-spin screening. In order to enforce 
this condition, we take a flat conduction-electron density of states of 
bandwidth 2~eV (the chemical potential is zero), but we assume that only 
the conduction electrons with energy $-D \leq \epsilon \leq D$ are coupled to 
the local degrees of freedom. Consequently, we perform the NRG procedure only 
on these electrons, which amounts to assume an effective bandwidth  
$2D \sim 2\times 10^{-2}$~eV, yet with a flat density of states equal to the original 
one, namely $0.5$~eV$^{-1}=0.5\times 10^{-2}/D$. 
Using the attempt frequency $D$ as our energy unit, the net result in the 
Wilson chain~\cite{wilson} is a renormalization of 
\[
V_{e(o)} \rightarrow \sqrt{\fract{D}{D_0}}\,V_{e(o)} = \fract{V_{e(o)}}{\sqrt{D\,D_0}}\,D = 2,
\]
which keeps the $d$-level hybridization width at the chemical potential invariant, 
while $V_x$ and $\Delta_x$ rescale trivially into themselves:
\[
V_x \rightarrow \fract{V_x}{D}\,D,\qquad \Delta_x\rightarrow \fract{\Delta_x}{D}\,D,
\]
implying $V_x\simeq \Delta_x\sim 10^{-2}~V_e$,~\cite{nota-xieta} the values we assume throughout. 
Moreover, to better identify each state of the spectrum, in the numerical calculations 
we implemented the spin SU(2) symmetry, the charge U(1) symmetry and the discrete parity 
defined by Eq.~\eqn{new-parity}.

Following the discussion of Sec. II, we ran NRG calculations for the two different 
implementations of the electron charge assisted tunnelling, i.e. case (i), in which Eq.~\eqn{condition} 
holds with $\eta=0$ and $\xi=1$, and  case (ii) with $\eta=\xi=0$. 

\begin{figure}[!htp]
\includegraphics*[width=\linewidth]{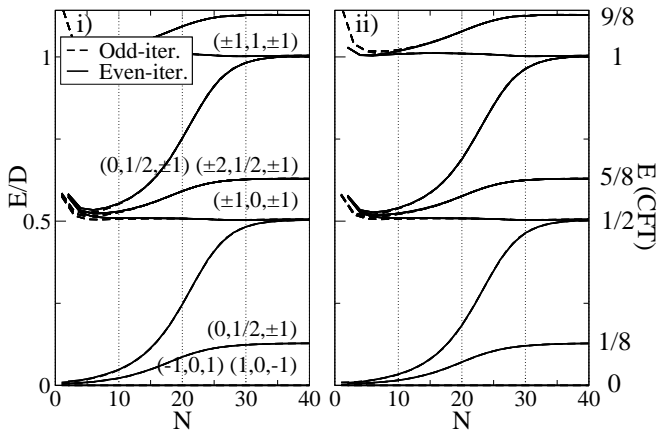}
\caption{NRG-flow of lowest eigenvalues for the model in Eq.~\ref{Ham-2}, 
with $\Delta_x=0$. The case (i) is analysed 
in the left panel, the case (ii) in the right one.}
\label{TLSd}
\end{figure}

In Fig.\ref{TLSd} we show the NRG-flow for the Hamiltonian in Eq.\ref{Ham-2} for both cases 
(i) and (ii) above with $\Delta_x=0$ and with $V_e=V_o=2$ and $V_x=10^{-2}~V_e$.  
The energy of the lowest lying eigenvalues are plotted as a function of the number 
$N$ of NRG iterations corresponding to an energy (temperature) scale $\omega_N = D \Lambda^{-N/2}$ 
where $\Lambda$ is the Wilson discretization parameter (we henceforth set $\Lambda=2$).  
At large $N$, the spacing between the levels, their degeneracy and the disappearance of  
any difference bewteen even and odd iterations $N$ (see for instance Ref.~\onlinecite{pangcox}) 
is typical of a 2CK. These results are summarized in Tab.\ref{tabTLSd}, and are 
consistent with the  conformal field theory prediction~\cite{ludwigaffl1,ludwigaffl2} for the 2CK.

The numerical results clearly show that, whatever the form of electron assisted 
tunneling, the system has a 2CK behavior at low temperatures. The Kondo temperature 
$T_K$ is conventionally estimated as  $D~ \Lambda^{-(N_c-1)/2}$, with $N_c$ the 
NRG-iteration at which e.g. the first excited state is 10\% off its 
asymptotic value.~\cite{pangcox,onzarand} We find that, while cases (i) and (ii) 
have roughly the same $N_c$, the latter is strongly influenced by $V_o/V_e$. 
In particular $V_o/V_e\simeq 1$, namely $\gamma\simeq 1$, is an optimal choice 
that minimizes $N_c\sim 25$, consistently with the previous analysis, and corresponds to a 
temperature of few hundredths of a Kelvin. Remarkably, even and odd iterations
are hardly distinguishable after very few iterations. That seems to be a property of 
the 2CK model right at its fixed point -- the fixed point with the highest 
$T_K\sim D$,~\cite{pangcox,onzarand} -- which would imply that the above estimate of 
$T_K$ is a strong underestimation of the real one. However, we cannot exclude 
the possibility that the even-odd collapse of the energy levels might simply indicate 
a preliminary crossover to a regime where the effects of $V_e$ and $V_o$ are fully 
established while those of $V_x$ are still negligible.

%
\begin{table}[!htp]
\begin{tabular}{|c|c||c|c|c||c|}\hline  
 E(CFT)&E(NRG) & $Q$  & $S$  & 	  $P$  &     deg     \\ \hline\hline
 0   & 0.0000        &  -1   & 0      & -1    & 1      \\ \hline
 0   & 0.0000       &   1   & 0      & 1&       1      \\ \hline
1/8 &0.1246 &   0   & 1/2   &$\pm$ 1     & 4      \\ \hline
1/2 &0.4999 &  +1   & 0       &-1    & 2  \\ \hline
1/2 &0.4999 &  -1   & 0        &+1   & 2  \\ \hline
1/2 &0.4999 &  +1   & 1       &+1    & 3      \\ \hline
1/2 &0.4999 &  -1   & 1       &-1    & 3      \\ \hline
5/8 &0.6290 &   0   & 1/2  &$\pm$1     & 4  \\ \hline
5/8 &0.6290 & $\pm$2& 1/2   &$\pm$1     & 8      \\ \hline
 1  &1.0230  &  -1   & 1       & +1    & 6  \\ \hline
 1  &1.0230  &   1   & 1       &-1    & 6  \\ \hline
\end{tabular} 
\caption{Lowest energy NRG spectrum of the Hamiltonian $H$ of \ref{Ham-2} for $\Delta_x=0$. 
The energies $E(NRG)$ are given in units of the fundamental level spacing 
and compared with the conformal field theory prediction $E(CFT)$. For each eigenvalue 
we indicate its degeneracy (deg) together with its quantum numbers 
$Q$ (charge), $S$ (spin) and parity $P$ defined in Eq.~\eqn{new-parity}.  
}
\label{tabTLSd} 
\end{table}

We note that, although the level spacings and degeneracies are those of the 
conventional 2CK model Eq.~\eqn{Ham-2CK}, the quantum numbers of each eigenvalue 
differ substantially from that model. In the flavour 2CK model, labeling 
states with $Q$, $S$ and the flavour $T$ (see Eq.~\eqn{flavour}), one expects 
the lowest energy spectrum of Table~\ref{tab3}. This spectrum is determined within 
conformal field theory~\cite{ludwigaffl1,ludwigaffl2} by so-called fusion of the 
free-electron spectrum, to the left in Tab.~\ref{tab3}, with the flavour primary field with $T=1/2$. 

\begin{table}[!htb]
\begin{tabular}{|c||c|c|c||c|}\hline  
 E(CFT)  &    $Q$  & $S$          & $T$    &deg   \\ \hline\hline
 0  &     0  & 0     & 0  & 1    \\ \hline
1/2 &       $\pm 1$ & 1/2  & 1/2    & 4    \\ \hline
1 &    0  & 1    & 1  & 9    \\ \hline
1 & $\pm$2 & 1    & 0  & 6    \\ \hline
1 & $\pm$2 & 0  & 1    & 6   \\ \hline
\end{tabular}~~

\begin{tabular}{|c||c|c|c||c|}\hline  
 E(CFT)  &    Q  & S          & T    &deg   \\ \hline\hline
 0  &     0  & 0     & 1/2  & 2    \\ \hline
1/8 &       $\pm 1$ & 1/2  & 0    & 2    \\ \hline
1/2 &    0  & 1    & 1/2  & 6    \\ \hline
1/2 & $\pm$2& 0    & 1/2  & 4    \\ \hline
5/8 & $\pm$1 & 1/2  & 1    & 12   \\ \hline
 1  & $\pm$2 & 1    & 1/2  & 12   \\ \hline
 \end{tabular}

\caption{Lowest energy spectrum of the 2CK model \eqn{Ham-2CK} for $V_a=0$, left table, 
and $V_a\not=0$, right table, as expected by conformal field theory.}
\label{tab3}
\end{table}

By contrast, we found that the NRG spectrum that we actually find, Tab.~\ref{tabTLSd}, can be 
obtained starting from the 2CK one in Tab.~\ref{tab3} in the following way.

\begin{itemize}
\item[(1)] First we decompose the flavour SU(2) $\rightarrow$ U(1)$\times$Z$_2$, where U(1) stands for the free bosonic 
theory that represents the $z$-component of the flavour field, and Z$_2$ is an Ising 
conformal field theory (see \S 18.5 in Ref.~\onlinecite{CFT}). This decomposition leads to the spectrum in 
Tab.~\ref{tab3-1}.~\cite{nota-CFT}

\begin{table}[!htb]
\begin{tabular}{|c||c|c|c|c|}\hline
E(CFT) &  $Q$      &  $S$  & $T_z$  & Z$_2$ \\ \hline\hline 
0      &  0      &  0  & $\pm$1 & $\sigma$    \\ \hline
1/8    & $\pm$ 1 & 1/2 & 0      & $I$         \\ \hline
1/2    &  0      &  1  & $\pm$1 & $\sigma$    \\ \hline      
1/2    &  $\pm$2 &  0  & $\pm$1 & $\sigma$    \\ \hline 
5/8    & $\pm$1  & 1/2 & 0      & $\epsilon$  \\ \hline
5/8    & $\pm$1  & 1/2 & $\pm$2 & $I$         \\ \hline
1      &  $\pm$2 &  1  & $\pm$1 & $\sigma$    \\ \hline 
9/8    & $\pm$ 1 & 1/2 & $\pm$2 &  $\epsilon$ \\ \hline
\end{tabular}

\caption{Lowest energy spectrum of the 2CK model upon decomposing the flavour 
SU(2) into U(1)$\times$Z$_2$. T$_z$ is the quantum number that defines the U(1) theory, 
while Z$_2$ corresponds to the coset theory, which is an Ising one. 
}
\label{tab3-1}
\end{table}
\item[(2)] Next we shift the charge $Q$ and $z$-component of the flavour 
$T_z$ by +1.~\cite{nota-CFT} This corresponds to the fact that 
the even chain has one more site. In this way we obtain the spectrum 
in Tab.~\ref{tab3-2} which coincides with that one in 
Tab.~\ref{tabTLSd}, including the degeneracy of each eigenvalue. 

\begin{table}[!htb]
\begin{tabular}{|c||c|c|c|c|}\hline
E(CFT) &  $Q$      &  $S$  & $T_z$  & Z$_2$ \\ \hline\hline 
0      &  $\pm$1 &  0  & 0 & $\sigma$    \\ \hline
1/8    &  0      & 1/2 & $\pm$1      & $I$         \\ \hline
1/2    & $\pm$1  & 1   &  0          & $\sigma$     \\ \hline
1/2      &  $\pm$1 &  0  & $\pm$2    & $\sigma$    \\ \hline
5/8    &  $\pm$2      & 1/2 & $\pm$1      & $I$         \\ \hline
5/8    &  0      & 1/2 & $\pm$1      & $\epsilon$   \\ \hline
1    & $\pm$1  & 1   &  $\pm$2          & $\sigma$     \\ \hline
9/8    &  $\pm$2      & 1/2 & $\pm$1      & $\epsilon$    \\ \hline
\end{tabular}
\caption{Lowest energy spectrum obtained from the one in Tab.~\ref{tab3-1} upon shifting $Q$ and $T_z$ by +1.}
\label{tab3-2}
\end{table}
\end{itemize}

We note that, if we recombine the charge U(1) with the Ising to form an isospin (charge) SU(2) theory, 
the spectrum becomes equal to the conventional 2CK one in Tab.~\ref{tab3} with the role 
of $Q$ played by $T_z$ and that of $T$ played by the isospin. In other words, it seems that, 
although the original model is not invariant under isospin SU(2) symmetry, the fixed  
point does in fact recover that symmetry.      
This unexpected result is confirmed by the spectrum calculated during the renormalization group procedure. 
Indeed, after very few iterations, the ground state becomes and stays for all $N>1$ doubly degenerate 
with quantum numbers $(Q,S)=(+1,0),(-1,0)$. 

The above observation also clarifies why the charge degrees of freedom play an important 
role once $\Delta_x$ is turned on.  
As said, a finite $\Delta_x$ is equivalent in the 2CK language to a magnetic field 
on the impurity site, which is known to be a relevant simmetry breaking perturbation 
destroying the anomalous 2CK behavior.~\cite{A&L+Cox} Indeed, we find that, as soon 
as $\Delta_x\not=0$, the spectrum flows to a Fermi-liquid one that 
can be interpreted as independent even and odd electron channels suffering different 
phase shifts $\delta_e$ and $\delta_o$. 

\begin{figure}[!htp]
        \includegraphics*[width=8cm]{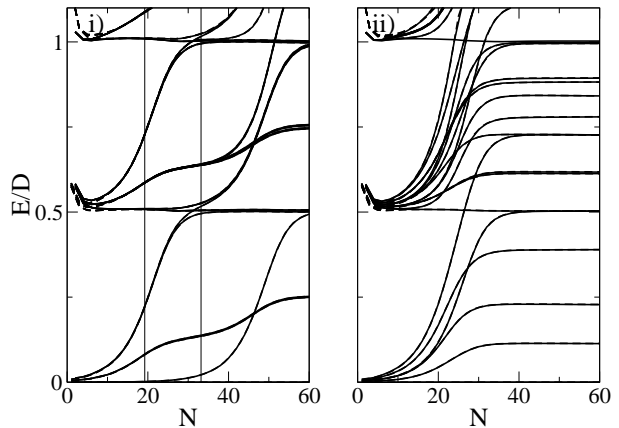}
\caption{NRG flow of the lowest eigenvalues for the model in Eq.\ref{Ham-2} with  
$\Delta_x=10^{-4}V_x$. Case (i) left panel, case (ii) right panel. Even and odd iterations correspond to 
solid and dashed lines, respectively.}
\label{spettro-delta}
\end{figure}

\begin{figure}[!htp]
        \includegraphics*[width=8cm]{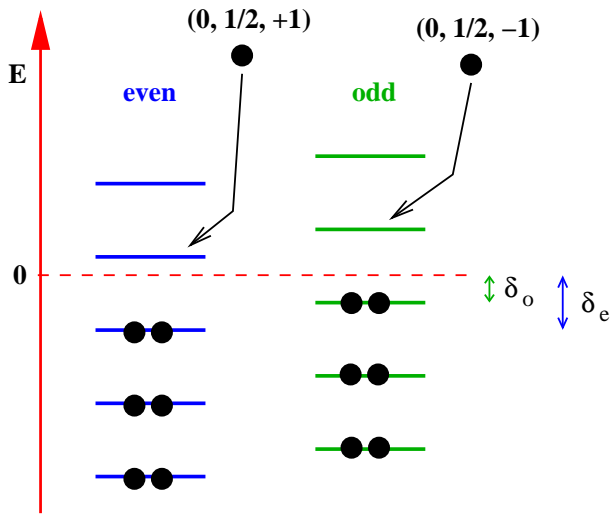}
\caption{Graphical representation of the Fermi-liquid spectrum for $\Delta_x\neq 0$. 
The even(odd) single particle energy levels 
are equidistant, however the even spectrum is shifted with respect to the odd one. 
The ground state is obtained by filling 
each level below the chemical potential, $E=0$ in the figure, and has quantum 
numbers $(Q,S,P)=(-1,0,+1)$. All possible 
excitations can be generated by combining single-particle excitations. 
We show for instance the two lowest energy excitations 
that amounts to adding one electron, either even, $(0,1/2,+1)$, or odd, 
$(0,1/2,-1)$, which we use to evaluate the phase shift 
difference $\delta_e-\delta_o$. 
}
\label{levels}
\end{figure}

In Fig.~\ref{spettro-delta} we show the NRG flow of the 
low energy spectrum for $\Delta_x=10^{-4} V_x$ for the two cases (i) and (ii). 
The asymptotic spectrum can be straightforwardly interpreted using the 
single-particle spectrum of Fig.~\ref{levels} and combining all possible single-particle excitations. 
In particular, we find that, for very small $\Delta_x\ll V_x$, $\delta_e-\delta_o=0$ 
for case (i) and $\delta_e-\delta_o=\pi/4$ for case (ii). A difference 
between the two cases is apparent also in the way their approach to the 
asymptotic behavior. In fact, for the same values of $\Delta_x\ll V_x$,  
a crossover region with a $\Delta_x=0$-spectrum is still visible in case (i) 
but not at all in case (ii), see Fig.~\ref{spettro-delta}. 
This different low energy behavior has its counterpart on the conductance behavior, 
as will be discussed in the next section.

We conclude this part by emphasizing that for realistic $\Delta_x\simeq V_x$ no 
crossover is visible in the spectrum, which might suggest the absence of any 
intermediate temperature regime dominated by the singular behavior of the 
2CK fixed point. However, this statement should be taken with caution, since, 
as discussed above, the flow, even at $\Delta_x=0$, is quite atypical and does not 
allow for a precise determination of $T_K$. Indeed, for $\Delta_x\not = 0$, it 
remains true that the spectra of even and odd iterations collapse very fast. However, 
unlike the case $\Delta_x=0$, the levels at even and odd iterations with equal 
energy have opposite charge $Q$ and parity $P$. This compares well with the role 
of a local magnetic field in the 2CK at its fixed point: 
levels at even and odd iterations with opposite spin quantum number $S_z$ collapse.~\cite{pangcox}
Therefore, although we tend to believe that the above estimate of $T_K\sim 10^{-4}~D$ is correct, 
we cannot exclude that the actual value could be much larger.

%
%

\section{Conductance}

We mentioned earlier that the zero-bias conductance in the 2CK state, $\Delta_x=0$, 
is zero because the scattering matrix of both the even and the odd channels are zero~\cite{A&L-green}.
For finite $\Delta_x$, the recovery of Fermi-liquid behavior allows us to estimate 
the conductance by the difference $\delta_e-\delta_o$, see Eq.~\eqn{G-e-o}, which 
can be extracted by the spectrum, for instance by calculating the energy difference 
between the two lowest energy states with $(Q,S,P)=(0,1/2,1)$ and $(0,1/2,-1)$ 
in units of the level spacing: 

\[
\delta_e-\delta_o=\pi \left( E_{(0,1/2,1)}- E_{(0,1/2,-1)}\right).
\] 

These two energies correspond to the cost of adding an even electron, 
$(Q,S,P)=(0,1/2,1)$, or an odd one, $(Q,S,P)=(0,1/2,-1)$, 
to the ground state, which has quantum numbers $(-1,0,1)$, see Fig.~\ref{levels}.   

\begin{figure}[!htp]
        \includegraphics*[width=\linewidth]{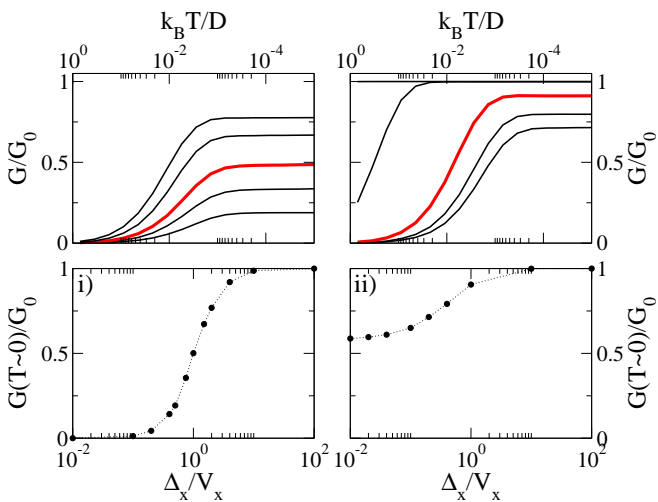}
\caption{Top panels. Conductance in units of the conductance quantum $G_0$ as a 
function of temperature for different values of $\Delta_x/V_x\sim 1$  for case (i) 
(left panel), and (ii) (right panel). Bottom panels: zero temperature conductance as a function 
of the ratio $\Delta_x/V_x$ for model (i) (left panel) and  (ii) (right panel).}
\label{conductance}
\end{figure}

We calculate this phase shift, hence the zero-bias conductance, as a function of the temperature $T$ 
(extracted from the NRG iterations), for different values of the ratio $\Delta_x/V_x$. 
We note however that, while we are quite confident about the values at low temperatures, 
those at high temperatures must be taken with caution since the spectrum is still 
far from a Fermi liquid one.   
The results are shown in the top panels of Fig.~\ref{conductance} for case (i) and case (ii) and 
realistic values of $\Delta_x/V_x\sim 1$ (red-bold curves).  
In both cases there is a significant thermal crossover with very small conductance 
before the asymptotic low temperature regime is reached. At zero temperature, 
the conductance is zero if $\Delta_x=0$. However, as soon as an infinitesimal 
$\Delta_x$ is turned on, the zero temperature conductance stays 0 in case (i) but jumps to $G_0/2$ 
in case (ii), see bottom panels of Fig.\ref{conductance}.
For realistic values of $\Delta_x\simeq V_x$, the zero temperature conductance 
is in all cases finite, $G\sim 0.5\div 0.9~G_0$, and smaller then the unitary value.

\section{Discussion and Conclusions} 

In summary, we have discussed the influence in the transport across a bridge atom 
of its quantum mechanical center-of-mass motion, whose dynamics in the double well  
case we have approximated by that of a two-level system~\cite{zawadowski}. In 
this regime, the two equilibrium positions of the bridge atom play the role of 
a pseudo-spin, whose dynamics is influenced by the electron hopping from the 
contacts into its valence orbital. This realizes effectively the same physics 
of a magnetic atom or a quantum dot bridging between two leads, the role of 
spin played by the position of the atom and the real spin playing the role 
of an additional flavour index. It is speculated that this hypothetical situation 
might be applicable to a metal break-junction caught right at the
breaking point, when the central atom bridging the two contacts develops,
although for a very short time interval, a double-well potential before collapsing 
finally onto one of the two. 

We find that, as long as the atom can tunnel between the two contacts, the 
zero-bias conductance at zero temperature is finite, although smaller than 
its value in the solid metal-metal nanocontact, with a single well for the
bridge atom. This finite conductance seems at variance with the earlier result
by Al-Hassanieh {\sl et al.}~\cite{Dagotto}, according to which the zero-temperature 
conductance at resonance should vanish at zero bias when the center-of-mass 
motion modulates the hopping amplitude into the leads. The discrepancy might 
be due to our two-level-system approximation or, more likely, to the different low-energy 
accuracy of NRG with respect to the numerical technique employed by Al-Hassanieh {\sl et al.}~\cite{Dagotto}. 
Indeed, we find that the finite-temperature conductance, which should correspond 
to the effective zero temperature value obtained with less low-energy 
accuracy, decreases quite rapidly towards zero with increasing temperature. 

In the limiting (and unrealistic) case of a vanishing spontaneous tunneling, $\Delta_x=0$, 
in spite of a finite assisted one, $V_x\not = 0$, the model displays a 
two-channel Kondo behavior, again with vanishing zero-temperature conductance. 
For finite $\Delta_x\ll V_x$, the zero temperature conductance is found 
either to remain zero or to jump to 1/2 of the unitary limit (the conductance quantum), 
$G=0.5 G_0=e^2/h$, depending on the form of the assisted tunneling. On the contrary, for realistic values 
of $\Delta_x\simeq V_x$, the conductance is always finite, $G\sim 0.5\div 0.9~G_0$.  

A critical aspect of the model is that, with the realistic parameters used, 
distinct signatures of the  two-level system dynamics could be hard to observe at 
temperatures around 4 K commonly used in metal break-junction experiments.~\cite{break_junction}
Even harder could be the detection of possible manifestations of two-channel-Kondo anomalies.
Cooling to lower temperature would offer the possibility to observe these effects. 
Time resolved conductance experiments could show
the tunneling regime as a transient just before breaking
and a coherent Kondo-like regime could be reached for 
light-mass shuttling-centers. For instance, hydrogen atoms or molecules
moving onto and into mechanically controllable break junctions (see e.g.  
Refs.~ \onlinecite{unpodiroba,kiguchi,csonka2}). In that case, the 
conductance plateaus found below the unitary limit, could be ascribable
to the two-level system dynamics, similarly to that found in our model, 
see Fig.~\ref{conductance}, now shifted to higher temperature scales.  
A possible realization could be a metal contact bridged by a malone aldehyde 
molecule, where a hydrogen bond is known to shuttle quantum mechanically between two equivalent positions
\cite{aldeide}.

Acknowledgments -- We are grateful to B. L. Altshuler, A. Tagliacozzo, D. Ugarte
and A. Zawadowski for enlightening discussions.  This research was
partially supported by MIUR-PRIN 2006022847. 
P.L. acknowledges finantial  support from CNR-INFM within 
ESF Eurocores Programme FoNE (Contract No.  ERAS-CT-2003- 980409).


\begin{thebibliography}{37}
\expandafter\ifx\csname natexlab\endcsname\relax\def\natexlab#1{#1}\fi
\expandafter\ifx\csname bibnamefont\endcsname\relax
  \def\bibnamefont#1{#1}\fi
\expandafter\ifx\csname bibfnamefont\endcsname\relax
  \def\bibfnamefont#1{#1}\fi
\expandafter\ifx\csname citenamefont\endcsname\relax
  \def\citenamefont#1{#1}\fi
\expandafter\ifx\csname url\endcsname\relax
  \def\url#1{\texttt{#1}}\fi
\expandafter\ifx\csname urlprefix\endcsname\relax\def\urlprefix{URL }\fi
\providecommand{\bibinfo}[2]{#2}
\providecommand{\eprint}[2][]{\url{#2}}

\bibitem[{\citenamefont{Galperin et~al.}(2007)\citenamefont{Galperin, Ratner,
  and Nitzan}}]{galperin}
\bibinfo{author}{\bibfnamefont{M.}~\bibnamefont{Galperin}},
  \bibinfo{author}{\bibfnamefont{M.~A.} \bibnamefont{Ratner}},
  \bibnamefont{and} \bibinfo{author}{\bibfnamefont{A.}~\bibnamefont{Nitzan}},
  \bibinfo{journal}{Journal of Physics: Condensed Matter}
  \textbf{\bibinfo{volume}{19}}, \bibinfo{pages}{103201 (81pp)}
  (\bibinfo{year}{2007}).

\bibitem[{\citenamefont{Al-Hassanieh et~al.}(2005)\citenamefont{Al-Hassanieh,
  Busser, Martins, and Dagotto}}]{Dagotto}
\bibinfo{author}{\bibfnamefont{K.~A.} \bibnamefont{Al-Hassanieh}},
  \bibinfo{author}{\bibfnamefont{C.~A.} \bibnamefont{Busser}},
  \bibinfo{author}{\bibfnamefont{G.~B.} \bibnamefont{Martins}},
  \bibnamefont{and} \bibinfo{author}{\bibfnamefont{E.}~\bibnamefont{Dagotto}},
  \bibinfo{journal}{Phys. Rev. Lett.} \textbf{\bibinfo{volume}{95}},
  \bibinfo{eid}{256807} (\bibinfo{year}{2005}).

\bibitem[{\citenamefont{Mravlje et~al.}(2006)\citenamefont{Mravlje, Ramsak, and
  Rejec}}]{Mravlje}
\bibinfo{author}{\bibfnamefont{J.}~\bibnamefont{Mravlje}},
  \bibinfo{author}{\bibfnamefont{A.}~\bibnamefont{Ramsak}}, \bibnamefont{and}
  \bibinfo{author}{\bibfnamefont{T.}~\bibnamefont{Rejec}},
  \bibinfo{journal}{Phys. Rev. B }
  \textbf{\bibinfo{volume}{74}}, \bibinfo{eid}{205320}
  (\bibinfo{year}{2006}).

\bibitem[{\citenamefont{Agrait et~al.}(2003)\citenamefont{Agrait, Yeyati, and
  van Ruitenbeek}}]{break_junction}
\bibinfo{author}{\bibfnamefont{N.}~\bibnamefont{Agrait}},
  \bibinfo{author}{\bibfnamefont{A.~L.~L.} \bibnamefont{Yeyati}},
  \bibnamefont{and} \bibinfo{author}{\bibfnamefont{J.~M.} \bibnamefont{van
  Ruitenbeek}}, \bibinfo{journal}{Phys. Rep.} \textbf{\bibinfo{volume}{377}},
  \bibinfo{eid}{81} (\bibinfo{year}{2003}).

\bibitem[{\citenamefont{Landauer}(1957)}]{LB1}
\bibinfo{author}{\bibfnamefont{R.}~\bibnamefont{Landauer}},
  \bibinfo{journal}{IBM J. Res. Dev.} \textbf{\bibinfo{volume}{1}},
  \bibinfo{pages}{223} (\bibinfo{year}{1957}).

\bibitem[{\citenamefont{Buttiker}(1988)}]{LB2}
\bibinfo{author}{\bibfnamefont{M.}~\bibnamefont{Buttiker}},
  \bibinfo{journal}{IBM J. Res. Dev.} \textbf{\bibinfo{volume}{32}},
  \bibinfo{pages}{317} (\bibinfo{year}{1988}).

\bibitem[{\citenamefont{Vladar and Zawadowski}(1983)}]{zawadowski}
\bibinfo{author}{\bibfnamefont{K.}~\bibnamefont{Vladar}} \bibnamefont{and}
  \bibinfo{author}{\bibfnamefont{A.}~\bibnamefont{Zawadowski}},
  \bibinfo{journal}{Phys. Rev. B} \textbf{\bibinfo{volume}{28}},
  \bibinfo{pages}{1564} (\bibinfo{year}{1983}).

\bibitem[{\citenamefont{{Nozi\`eres} and Blandin}(1980)}]{nozieres}
\bibinfo{author}{\bibfnamefont{P.}~\bibnamefont{{Nozi\`eres}}}
  \bibnamefont{and} \bibinfo{author}{\bibfnamefont{A.}~\bibnamefont{Blandin}},
  \bibinfo{journal}{J. Phys. (Paris)} \textbf{\bibinfo{volume}{41}},
  \bibinfo{pages}{193} (\bibinfo{year}{1980}).

\bibitem[{\citenamefont{Ralph et~al.}(1994)\citenamefont{Ralph, Ludwig, {von
  Delft}, and Buhrman}}]{ralph}
\bibinfo{author}{\bibfnamefont{D.~C.} \bibnamefont{Ralph}},
  \bibinfo{author}{\bibfnamefont{A.~W.~W.} \bibnamefont{Ludwig}},
  \bibinfo{author}{\bibfnamefont{J.}~\bibnamefont{{von Delft}}},
  \bibnamefont{and} \bibinfo{author}{\bibfnamefont{R.~A.}
  \bibnamefont{Buhrman}}, \bibinfo{journal}{Phys. Rev. Lett.}
  \textbf{\bibinfo{volume}{72}}, \bibinfo{pages}{1064} (\bibinfo{year}{1994}).

\bibitem[{\citenamefont{Halbritter et~al.}(2000)\citenamefont{Halbritter,
  Kolesnychenko, {Mih\'aly}, Shklyarevskii, and van Kempen}}]{halbritter}
\bibinfo{author}{\bibfnamefont{A.}~\bibnamefont{Halbritter}},
  \bibinfo{author}{\bibfnamefont{O.~Y.} \bibnamefont{Kolesnychenko}},
  \bibinfo{author}{\bibfnamefont{G.}~\bibnamefont{{Mih\'aly}}},
  \bibinfo{author}{\bibfnamefont{O.~I.} \bibnamefont{Shklyarevskii}},
  \bibnamefont{and} \bibinfo{author}{\bibfnamefont{H.}~\bibnamefont{van
  Kempen}}, \bibinfo{journal}{Phys. Rev. B} \textbf{\bibinfo{volume}{61}},
  \bibinfo{pages}{5846} (\bibinfo{year}{2000}).

\bibitem[{\citenamefont{Cichorek et~al.}(2005)\citenamefont{Cichorek, Sanchez,
  Gegenwart, Weickert, Wojakowski, Henkie, Auffermann, Paschen, Kniep, and
  Steglich}}]{cichorek}
\bibinfo{author}{\bibfnamefont{T.}~\bibnamefont{Cichorek}},
  \bibinfo{author}{\bibfnamefont{A.}~\bibnamefont{Sanchez}},
  \bibinfo{author}{\bibfnamefont{P.}~\bibnamefont{Gegenwart}},
  \bibinfo{author}{\bibfnamefont{F.}~\bibnamefont{Weickert}},
  \bibinfo{author}{\bibfnamefont{A.}~\bibnamefont{Wojakowski}},
  \bibinfo{author}{\bibfnamefont{Z.}~\bibnamefont{Henkie}},
  \bibinfo{author}{\bibfnamefont{G.}~\bibnamefont{Auffermann}},
  \bibinfo{author}{\bibfnamefont{S.}~\bibnamefont{Paschen}},
  \bibinfo{author}{\bibfnamefont{R.}~\bibnamefont{Kniep}}, \bibnamefont{and}
  \bibinfo{author}{\bibfnamefont{F.}~\bibnamefont{Steglich}},
  \bibinfo{journal}{Phys. Rev. Lett.} \textbf{\bibinfo{volume}{94}},
  \bibinfo{pages}{236603} (\bibinfo{year}{2005}).

\bibitem[{\citenamefont{Zawadowski et~al.}(1999)\citenamefont{Zawadowski, {von
  Delft}, and Ralph}}]{zawadowski2}
\bibinfo{author}{\bibfnamefont{A.}~\bibnamefont{Zawadowski}},
  \bibinfo{author}{\bibfnamefont{J.}~\bibnamefont{{von Delft}}},
  \bibnamefont{and} \bibinfo{author}{\bibfnamefont{D.~C.} \bibnamefont{Ralph}},
  \bibinfo{journal}{Phys. Rev. Lett.} \textbf{\bibinfo{volume}{83}},
  \bibinfo{pages}{2632} (\bibinfo{year}{1999}).

\bibitem[{\citenamefont{Kolesnychenko et~al.}(2002)\citenamefont{Kolesnychenko,
  {de Kort}, Katsnelson, Lichtenstein, and van Kempen}}]{kolesnychenko}
\bibinfo{author}{\bibfnamefont{O.~Y.} \bibnamefont{Kolesnychenko}},
  \bibinfo{author}{\bibfnamefont{R.}~\bibnamefont{{de Kort}}},
  \bibinfo{author}{\bibfnamefont{M.~I.} \bibnamefont{Katsnelson}},
  \bibinfo{author}{\bibfnamefont{A.~I.} \bibnamefont{Lichtenstein}},
  \bibnamefont{and} \bibinfo{author}{\bibfnamefont{H.}~\bibnamefont{van
  Kempen}}, \bibinfo{journal}{Nature} \textbf{\bibinfo{volume}{415}},
  \bibinfo{pages}{507} (\bibinfo{year}{2002}).

\bibitem[{\citenamefont{Cox and Zawadowski}(1998)}]{cox}
\bibinfo{author}{\bibfnamefont{D.~L.} \bibnamefont{Cox}} \bibnamefont{and}
  \bibinfo{author}{\bibfnamefont{A.}~\bibnamefont{Zawadowski}},
  \bibinfo{journal}{Adv. Phys.} \textbf{\bibinfo{volume}{47}},
  \bibinfo{pages}{599} (\bibinfo{year}{1998}).

\bibitem[{\citenamefont{Aleiner et~al.}(2001)\citenamefont{Aleiner, Altshuler,
  Galperin, and Shutenko}}]{altshuler}
\bibinfo{author}{\bibfnamefont{I.~L.} \bibnamefont{Aleiner}},
  \bibinfo{author}{\bibfnamefont{B.~L.} \bibnamefont{Altshuler}},
  \bibinfo{author}{\bibfnamefont{Y.~M.} \bibnamefont{Galperin}},
  \bibnamefont{and} \bibinfo{author}{\bibfnamefont{T.~A.}
  \bibnamefont{Shutenko}}, \bibinfo{journal}{Phys. Rev. Lett.}
  \textbf{\bibinfo{volume}{86}}, \bibinfo{pages}{2629} (\bibinfo{year}{2001}).

\bibitem[{\citenamefont{Wilson}(1975)}]{wilson}
\bibinfo{author}{\bibfnamefont{K.}~\bibnamefont{Wilson}},
  \bibinfo{journal}{Rev. Mod. Phys.} \textbf{\bibinfo{volume}{47}},
  \bibinfo{pages}{773} (\bibinfo{year}{1975}).

\bibitem[{\citenamefont{Zarand}(2005)}]{zarand}
\bibinfo{author}{\bibfnamefont{G.}~\bibnamefont{Zarand}},
  \bibinfo{journal}{Phys. Rev. B} \textbf{\bibinfo{volume}{72}},
  \bibinfo{pages}{245103} (\bibinfo{year}{2005}).

\bibitem[{\citenamefont{Kolf and Kroha}(2007)}]{onzarand}
\bibinfo{author}{\bibfnamefont{C.}~\bibnamefont{Kolf}} \bibnamefont{and}
  \bibinfo{author}{\bibfnamefont{J.}~\bibnamefont{Kroha}},
  \bibinfo{journal}{Phys. Rev. B}
  \textbf{\bibinfo{volume}{75}}, \bibinfo{eid}{045129}
  (\bibinfo{year}{2007}).

\bibitem[{\citenamefont{Datta}(1995)}]{datta}
\bibinfo{author}{\bibfnamefont{S.}~\bibnamefont{Datta}},
  \emph{\bibinfo{title}{Electronic Transport in Mesoscopic Systems}}
  (\bibinfo{publisher}{Cambridge Studies in Semiconductor Physics Series},
  \bibinfo{year}{1995}).

\bibitem[{\citenamefont{Yuval and Anderson}(1970)}]{Anderson&Yuval}
\bibinfo{author}{\bibfnamefont{G.}~\bibnamefont{Yuval}} \bibnamefont{and}
  \bibinfo{author}{\bibfnamefont{P.~W.} \bibnamefont{Anderson}},
  \bibinfo{journal}{Phys. Rev. B} \textbf{\bibinfo{volume}{1}},
  \bibinfo{pages}{1522} (\bibinfo{year}{1970}).

\bibitem[{\citenamefont{Hamann}(1970)}]{Hamann}
\bibinfo{author}{\bibfnamefont{D.~R.} \bibnamefont{Hamann}},
  \bibinfo{journal}{Phys. Rev. B} \textbf{\bibinfo{volume}{2}},
  \bibinfo{pages}{1373} (\bibinfo{year}{1970}).

\bibitem[{\citenamefont{Nozi\`eres and
  De~Dominicis}(1969)}]{Nozieres&DeDominicis}
\bibinfo{author}{\bibfnamefont{P.}~\bibnamefont{Nozi\`eres}} \bibnamefont{and}
  \bibinfo{author}{\bibfnamefont{C.~T.} \bibnamefont{De~Dominicis}},
  \bibinfo{journal}{Phys. Rev.} \textbf{\bibinfo{volume}{178}},
  \bibinfo{pages}{1097} (\bibinfo{year}{1969}).

\bibitem[{\citenamefont{Vlad\'ar et~al.}(1988)\citenamefont{Vlad\'ar,
  Zawadowski, and Zim\'anyi}}]{vladar}
\bibinfo{author}{\bibfnamefont{K.}~\bibnamefont{Vlad\'ar}},
  \bibinfo{author}{\bibfnamefont{A.}~\bibnamefont{Zawadowski}},
  \bibnamefont{and} \bibinfo{author}{\bibfnamefont{G.~T.}
  \bibnamefont{Zim\'anyi}}, \bibinfo{journal}{Phys. Rev. B}
  \textbf{\bibinfo{volume}{37}}, \bibinfo{pages}{2001} (\bibinfo{year}{1988}).

\bibitem[{\citenamefont{Fabrizio et~al.}(1995)\citenamefont{Fabrizio, Gogolin,
  and Nozi\`eres}}]{mio}
\bibinfo{author}{\bibfnamefont{M.}~\bibnamefont{Fabrizio}},
  \bibinfo{author}{\bibfnamefont{A.~O.} \bibnamefont{Gogolin}},
  \bibnamefont{and}
  \bibinfo{author}{\bibfnamefont{P.}~\bibnamefont{Nozi\`eres}},
  \bibinfo{journal}{Phys. Rev. B} \textbf{\bibinfo{volume}{51}},
  \bibinfo{pages}{16088} (\bibinfo{year}{1995}).

\bibitem[{\citenamefont{Emery and Kivelson}(1992)}]{Emery&Kivelson}
\bibinfo{author}{\bibfnamefont{V.~J.} \bibnamefont{Emery}} \bibnamefont{and}
  \bibinfo{author}{\bibfnamefont{S.}~\bibnamefont{Kivelson}},
  \bibinfo{journal}{Phys. Rev. B} \textbf{\bibinfo{volume}{46}},
  \bibinfo{pages}{10812} (\bibinfo{year}{1992}).

\bibitem[{\citenamefont{Affleck et~al.}(1992)\citenamefont{Affleck, Ludwig,
  Pang, and Cox}}]{A&L+Cox}
\bibinfo{author}{\bibfnamefont{I.}~\bibnamefont{Affleck}},
  \bibinfo{author}{\bibfnamefont{A.~W.~W.} \bibnamefont{Ludwig}},
  \bibinfo{author}{\bibfnamefont{H.-B.} \bibnamefont{Pang}}, \bibnamefont{and}
  \bibinfo{author}{\bibfnamefont{D.~L.} \bibnamefont{Cox}},
  \bibinfo{journal}{Phys. Rev. B} \textbf{\bibinfo{volume}{45}},
  \bibinfo{pages}{7918} (\bibinfo{year}{1992}).

\bibitem[{\citenamefont{Pang and Cox}(1991)}]{pangcox}
\bibinfo{author}{\bibfnamefont{H.~B.} \bibnamefont{Pang}} \bibnamefont{and}
  \bibinfo{author}{\bibfnamefont{D.~L.} \bibnamefont{Cox}},
  \bibinfo{journal}{Phys. Rev. B} \textbf{\bibinfo{volume}{44}},
  \bibinfo{pages}{9454} (\bibinfo{year}{1991}).

\bibitem[{\citenamefont{Affleck and Ludwig}(1993)}]{A&L-green}
\bibinfo{author}{\bibfnamefont{I.}~\bibnamefont{Affleck}} \bibnamefont{and}
  \bibinfo{author}{\bibfnamefont{A.~W.~W.} \bibnamefont{Ludwig}},
  \bibinfo{journal}{Phys. Rev. B} \textbf{\bibinfo{volume}{48}},
  \bibinfo{pages}{7297} (\bibinfo{year}{1993}).

\bibitem[{not({\natexlab{a}})}]{nota-xieta}
\bibinfo{note}{We note however that the original values of $\xi$ and $\eta$ in
  Eq.~\eqn{Ham-2} must be rescaled by $10^{-2}$. However, since their precise
  values are uncertain, we do not include such a rescaling explicitly.}

\bibitem[{\citenamefont{Affleck and Ludwig}(1991{\natexlab{a}})}]{ludwigaffl1}
\bibinfo{author}{\bibfnamefont{I.}~\bibnamefont{Affleck}} \bibnamefont{and}
  \bibinfo{author}{\bibfnamefont{A.~W.~W.} \bibnamefont{Ludwig}},
  \bibinfo{journal}{Nucl. Phys. B} \textbf{\bibinfo{volume}{352}},
  \bibinfo{pages}{849} (\bibinfo{year}{1991}{\natexlab{a}}).

\bibitem[{\citenamefont{Affleck and Ludwig}(1991{\natexlab{b}})}]{ludwigaffl2}
\bibinfo{author}{\bibfnamefont{I.}~\bibnamefont{Affleck}} \bibnamefont{and}
  \bibinfo{author}{\bibfnamefont{A.~W.~W.} \bibnamefont{Ludwig}},
  \bibinfo{journal}{Nucl. Phys. B} \textbf{\bibinfo{volume}{360}},
  \bibinfo{pages}{641} (\bibinfo{year}{1991}{\natexlab{b}}).

\bibitem[{\citenamefont{{Di Francesco} et~al.}(1997)\citenamefont{{Di
  Francesco}, Mathieu, and S\'en\'echal}}]{CFT}
\bibinfo{author}{\bibfnamefont{P.}~\bibnamefont{{Di Francesco}}},
  \bibinfo{author}{\bibfnamefont{P.}~\bibnamefont{Mathieu}}, \bibnamefont{and}
  \bibinfo{author}{\bibfnamefont{D.}~\bibnamefont{S\'en\'echal}},
  \emph{\bibinfo{title}{Conformal Field Theory}}
  (\bibinfo{publisher}{Springer-Verlag New York, Inc.}, \bibinfo{year}{1997}).

\bibitem[{not({\natexlab{b}})}]{nota-CFT}
\bibinfo{note}{Notice that, within conformal field theory~\cite{CFT}, the
  abelian sectors corresponding to the charge $Q$ and the flavour $T_z$ are
  labeled by an integer $m$ defined modulo 4, e.g. $m=-1,0,1,2$. Therefore
  $m=+2$ and $m=-2$ have to be identified; they correspond in conformal field
  theory to the same character although physically to two different states.
  Seemingly, if we shift up by one the quantum numbers, then $m=2$, which is
  the same as $m=-2$, transforms into $m=-1$.}

\bibitem[{\citenamefont{Csonka et~al.}(2003)\citenamefont{Csonka, Halbritter,
  {Mih\'aly}, Jurdik, Shklyarevskii, Speller, and van
  Kempen}}]{unpodiroba}
\bibinfo{author}{\bibfnamefont{S.}~\bibnamefont{Csonka}},
  \bibinfo{author}{\bibfnamefont{A.}~\bibnamefont{Halbritter}},
  \bibinfo{author}{\bibfnamefont{G.}~\bibnamefont{{Mih\'aly}}},
  \bibinfo{author}{\bibfnamefont{E.}~\bibnamefont{Jurdik}},
  \bibinfo{author}{\bibfnamefont{O.~I.} \bibnamefont{Shklyarevskii}},
  \bibinfo{author}{\bibfnamefont{S.}~\bibnamefont{Speller}}, \bibnamefont{and}
  \bibinfo{author}{\bibfnamefont{H.}~\bibnamefont{van Kempen}},
  \bibinfo{journal}{Phys. Rev. Lett.} \textbf{\bibinfo{volume}{90}},
  \bibinfo{pages}{116803} (\bibinfo{year}{2003}).

\bibitem[{\citenamefont{Kiguchi et~al.}(2006)\citenamefont{Kiguchi, Konishi,
  and Murakoshi}}]{kiguchi}
\bibinfo{author}{\bibfnamefont{M.}~\bibnamefont{Kiguchi}},
  \bibinfo{author}{\bibfnamefont{T.}~\bibnamefont{Konishi}}, \bibnamefont{and}
  \bibinfo{author}{\bibfnamefont{K.}~\bibnamefont{Murakoshi}},
  \bibinfo{journal}{Phys. Rev. B} \textbf{\bibinfo{volume}{73}},
  \bibinfo{pages}{125406} (\bibinfo{year}{2006}).

\bibitem[{\citenamefont{Csonka et~al.}(2006)\citenamefont{Csonka, Halbritter,
  and {Mih\'aly}}}]{csonka2}
\bibinfo{author}{\bibfnamefont{S.}~\bibnamefont{Csonka}},
  \bibinfo{author}{\bibfnamefont{A.}~\bibnamefont{Halbritter}},
  \bibnamefont{and}
  \bibinfo{author}{\bibfnamefont{G.}~\bibnamefont{{Mih\'aly}}},
  \bibinfo{journal}{Phys. Rev. B} \textbf{\bibinfo{volume}{73}},
  \bibinfo{pages}{075405} (\bibinfo{year}{2006}).

\bibitem[{\citenamefont{Bunker}(1979)}]{aldeide}
\bibinfo{author}{\bibfnamefont{P.~R.} \bibnamefont{Bunker}},
  \emph{\bibinfo{title}{Molecular Symmetry and Spectroscopy}}
  (\bibinfo{publisher}{Academic, New York}, \bibinfo{year}{1979}).

\end{thebibliography}

\end{document}